\documentclass[10pt,conference]{IEEEtran}
\IEEEoverridecommandlockouts

\usepackage[utf8]{inputenc}
\usepackage[T1]{fontenc}
\usepackage{microtype}
\usepackage{graphicx}
\usepackage{booktabs}   
\usepackage{url}
\usepackage{xcolor}
\usepackage{balance}

\title{Grid-STIX: A STIX 2.1-Compliant Cyber-Physical Security Ontology for Power Grid and Nuclear Energy Systems}

\author{%
\IEEEauthorblockN{Benjamin Blakely,
Daniel Karcz}
\IEEEauthorblockA{Strategic Security Sciences, Argonne National Laboratory\\
\texttt{bblakely, dkarcz}@anl.gov}
}

\begin{document}

\maketitle

\begin{abstract}
\noindent
Modern electrical power grids represent complex cyber-physical systems requiring specialized cybersecurity frameworks beyond traditional IT security models. Existing threat intelligence standards such as STIX 2.1 and MITRE ATT\&CK lack coverage for grid-specific assets, operational technology relationships, and cyber-physical interdependencies essential for power system security. We present Grid-STIX, a domain-specific extension of STIX 2.1 for electrical grid cybersecurity applications. Grid-STIX employs a modular architecture encompassing physical assets, operational technology components, cyber-physical relationships, and security policies that capture modern power systems including distributed energy resources, advanced metering infrastructure, and nuclear energy facilities. The framework provides threat modeling capabilities through systematic representation of attack patterns, supply chain risks, and cross-domain impact analysis while maintaining STIX 2.1 compliance. Grid-STIX includes modules for nuclear safeguards and non-proliferation verification, enabling cybersecurity modeling across conventional and nuclear energy sectors. The ontology supports Zero Trust enforcement through policy decision points and operational context integration. Our implementation includes validation pipelines, Python code generation, and visualizations. Use cases demonstrate applications including cross-utility threat intelligence sharing, supply chain risk assessment, and nuclear facility cybersecurity. Grid-STIX is available as an open-source framework to advance collaborative cybersecurity research across the electrical power sector.
\end{abstract}


\section{Introduction}
Modern electrical power grids represent archetypal cyber-physical systems (CPS) that integrate computing, communication, and control capabilities with physical infrastructure for electricity generation, transmission, and distribution ~\cite{humayedCyberPhysicalSystemsSecurity2017,yuSurveyCyberPhysical2023}. The tight coupling between cyber and physical power infrastructure creates unique security challenges demanding threat modeling approaches that capture cyber-physical interdependencies and operational constraints.

Existing cybersecurity standards such as Structured Threat Information Expression (STIX) and MITRE's Adversarial Tactics, Techniques, and Common Knowledge (ATT\&CK) for Industrial Control Systems (ICS) provide valuable primitives but lack coverage for grid-specific cyber-physical relationships, operational technology (OT) dependencies, and latency-constrained control paths essential for power system security ~\cite{jiangMITREATTCKApplications2025}. Grid-STIX addresses these gaps with a cyber-physical security ontology modeling grid assets, protocols, protection relationships, and supply-chain risk while maintaining STIX 2.1 compliance.

\paragraph*{Contributions}
\begin{itemize}
  \item A STIX 2.1–aligned cyber-physical security ontology for electrical grid systems with support for nuclear energy facilities and safeguards applications.
  \item A validation framework and automated code-generation pipeline producing STIX-compliant Python classes with visualizations for communication.
  \item Semantic foundations for Zero Trust enforcement in distributed energy resource networks and nuclear facility cybersecurity applications.
\end{itemize}

\section{Background and Related Work}

Grid-STIX is a domain-specific STIX 2.1 extension designed to address cybersecurity challenges in distributed energy resource (DER) systems and electrical power grid infrastructure~\cite{zografopoulosDistributedEnergyResources2023,liuEnhancingCyberResiliencyDERbased2024}. 

\subsection{Threat Intelligence Standards}

Modern threat intelligence sharing relies on common ontologies and protocols to allow information exchange and automated processing. The STIX/Trusted Automated eXchange of Intelligence Information (TAXII) framework from OASIS provides the foundation for standardized cyber threat intelligence exchange~\cite{oasisSTIX21,STIXBestPractices2022}. STIX 2.1 uses JSON to standardize cyber threat intelligence through 18 domain objects, 18 cyber-observable objects, relationship objects, and vocabulary sets, enabling automated threat intelligence sharing across government and industry organizations. The companion TAXII 2.1 standard enables automated distribution through RESTful APIs~\cite{oasisTAXII21}. Grid-STIX extends this framework to address the unique requirements of softwarized cyber-physical systems (CPS), particularly modern electrical grids with DER, advanced metering infrastructure (AMI), and software-defined control systems.

Current threat intelligence practices follow established intelligence lifecycles encompassing collection from open-source intelligence (OSINT) and proprietary sources, processing through normalization and enrichment, and dissemination via automated feeds integrated with security information and event management (SIEM) and security orchestration, automation and response (SOAR) platforms~\cite{gongCyberThreatIntelligence2021, mavroeidisCyberThreatIntelligence2017}. However, these frameworks exhibit fundamental limitations when applied to CPS and OT environments~\cite{banerjeeEnsuringSafetySecurity2012,wolfSafetySecurityCyberPhysical2018}.

Research shows that while MITRE ATT\&CK integration with STIX works for IT environments, it exhibits major gaps for power grid applications~\cite{jiangMITREATTCKApplications2025}. Key limitations include inadequate representation of physical processes, insufficient OT protocol support, and missing safety considerations crucial for power grids. The ICS-specific extension (ATT\&CK for ICS) helps remediate this, but needs integration with a broader ontology to maximally useful. Grid-STIX addresses these gaps through comprehensive cyber-physical security extensions detailed in subsequent sections.

\subsection{OT/CPS Security and Standards}

OT/CPS security in power grids involves multiple standards including IEC 62351 for communication protocol security and North American Electric Reliability Corporation (NERC) Critical Infrastructure Protection (CIP) for bulk electric system cybersecurity~\cite{ghiasiComprehensiveReviewCyberattacks2023,thesmartgridinteroperabilitypanel-smartgridcybersecuritycommitteeGuidelinesSmartGrid2014}. Industrial communication protocols present diverse security postures and implementation challenges that affect threat modeling approaches, which Grid-STIX addresses through comprehensive vulnerability representation.

\subsection{Prior Ontologies and Knowledge Graphs}

Security ontologies for ICS include the Unified Cybersecurity Ontology (UCO) for IT security, FireEye's OT-CSIO for OT, and the Vocabulary for Event Recording and Incident Sharing (VERIS) Community Database for incident classification, but gaps remain for power grid cybersecurity ~\cite{iannaconeDevelopingOntologyCyber2015,tefekSmartGridOntology2023}. Existing STIX 2.1 frameworks focus on traditional IT security rather than power system requirements~\cite{mavroeidisCyberThreatIntelligence2017}. 

Idaho National Laboratory's STIX for ICS Threat Intelligence Graph (STIG)~\cite{inlSTIG} provides STIX 2.1 extensions for generic ICS environments, focusing on manufacturing and process control systems. While STIG establishes foundational ICS threat modeling patterns, it lacks power system-specific constructs including protection relay coordination, cascading failure dependencies, distributed energy resource orchestration, and nuclear safeguards integration essential for electrical grid cybersecurity. Gaps persist in modeling power protection relationships, grid synchronization dependencies, and supply-chain risk representation for power systems~\cite{islamOntologybasedUserPrivacy2022,thesmartgridinteroperabilitypanel-smartgridcybersecuritycommitteeGuidelinesSmartGrid2014}.

The IEC Common Information Model (CIM)~\cite{CommonInformationModel,cimModelingGuide,schumilinOntologybasedNetworkModel2017a} provides standardized data models for power system operations and energy management but focuses on operational data exchange rather than cybersecurity threat intelligence. While CIM excels at modeling power system topology and operational state, it lacks threat-specific constructs including attack patterns, vulnerability tracking, and adversarial tactics essential for cybersecurity applications. Grid-STIX complements CIM by providing cybersecurity-focused extensions that bridge operational power system data with STIX-based threat intelligence.

Grid-STIX addresses these gaps as the first cybersecurity ontology designed for electrical power systems. It enables standardized threat intelligence sharing, automated security analysis integration, and bridges STIX-based intelligence with power operations data. This advancement integrates protection system modeling, AMI standardization, and supply-chain risk representation within a unified STIX 2.1-compliant framework for zero-trust deployments.

\section{Grid Cyber Threat Model}

Grid-STIX provides threat modeling capabilities for cyber-physical power systems, detailing unique attack vectors and demonstrating how Grid-STIX organizes threat information through structured STIX 2.1 extensions.

\subsection{Attack Vectors}

Electrical power grids present unique attack surfaces extending beyond traditional IT environments, requiring threat modeling approaches that account for cyber-physical interdependencies, real-time operational constraints, and cascading failure potential ~\cite{liuEnhancingCyberResiliencyDERbased2024,ghiasiComprehensiveReviewCyberattacks2023}.

\paragraph*{Coordinated Asset-Centric Attack Progression}
Grid threat actors employ coordinated attack strategies that exploit multiple entry points and follow asset-specific progression patterns. These campaigns combine network infiltration with social engineering to gain initial access, then leverage protocol vulnerabilities for lateral movement across operational technology networks. Attacks typically begin with edge device compromise (smart meters, field sensors) and progress through communication networks toward higher-value targets (substations, control centers).

Time-synchronized attacks exploit operational timing windows such as peak demand periods, planned maintenance, or emergency conditions when grid operators have reduced situational awareness and limited response options. DER ecosystems enable distributed attack coordination where compromised solar inverters, battery systems, and electric vehicle charging infrastructure can be orchestrated to create grid-wide instability across multiple administrative domains.

\paragraph*{Protocol Diversity and Attack Surface Expansion}
Modern grid environments employ diverse protocol ecosystems that create complex, heterogeneous attack surfaces spanning safety-critical and operational systems. Safety instrumentation and control systems utilize Triple Modular Redundancy (TMR) protocols, PROFIsafe, and GOOSE messaging within IEC 61850 frameworks, where compromise can directly impact protection system reliability and personnel safety. Non-safety industrial control systems deploy Modbus, OPC UA, Ethernet/IP, and DNP3 protocols with varying security capabilities and authentication mechanisms.

Protocol-specific vulnerabilities significantly expand attack surfaces across grid infrastructure ~\cite{obertRecommendationsTrustEncryption2019,johnsonCybersecurityDevices}. DNP3, used widely in North American utilities, lacks built-in security mechanisms and suffers from man-in-the-middle attacks and replay vulnerabilities. IEC 61850's GOOSE protocol operates without authentication over multicast Ethernet, making it vulnerable to injection attacks that can manipulate critical circuit breaker operations. Grid interface protocols enable wide-area coordination but introduce inter-utility trust dependencies and cross-domain attack vectors.

This protocol heterogeneity complicates security policy enforcement, vulnerability management, and incident response across operational domains. Attack campaigns can exploit protocol bridging and gateway vulnerabilities to traverse security boundaries between safety and non-safety systems, potentially escalating operational disruptions to safety-critical failures.

\paragraph*{Shared Responsibility and Physical Interconnection Dependencies}
Electrical grids present unique cybersecurity challenges through shared responsibility models where interconnected assets under disjoint ownership and control must maintain strict physical coordination. Unlike traditional IT systems, grid assets are coupled through fundamental electrical properties—voltage, frequency, and phase synchronization—that create unavoidable interdependencies across organizational boundaries. Compromised assets under one entity's control can directly impact grid stability for all interconnected participants through physical laws governing electrical power flow.

This shared responsibility extends to real-time operational dependencies where cybersecurity failures propagate through physical mechanisms. Protection system cyberattacks can cause protective relay miscoordination that cascades across utility boundaries, creating response coordination challenges.

\paragraph*{Cascading Impact Modeling}
Power system interdependencies enable cyber attacks to trigger cascading failures that extend far beyond initial compromise scope. Protection system manipulation can cause equipment misoperation leading to widespread blackouts. Market manipulation attacks can exploit real-time pricing mechanisms to create artificial scarcity or demand spikes. Load forecasting attacks can manipulate predictive models to cause resource allocation errors.

Grid-STIX threat modeling captures these cascading relationships to enable impact assessment and defensive prioritization. Attack patterns include temporal dependencies, prerequisite conditions, and amplification factors that determine how localized compromise can escalate to system-wide consequences.

\paragraph*{Zero Trust Attack Surface Analysis}
Traditional network perimeter security fails in modern grid environments where assets span multiple administrative domains, communication networks, and trust boundaries. Zero Trust threat modeling requires granular analysis of every asset, communication path, and trust relationship as potential attack vectors.

Grid-STIX models trust boundaries at multiple levels: device authentication, communication channel integrity, operational authority validation, and cross-domain policy enforcement. This enables attack surface analysis that accounts for trust erosion, privilege escalation, and lateral movement across heterogeneous grid infrastructure.

\subsection{Organizing Cyber Threat Information}

Grid-STIX addresses the complexity of power system threat modeling through structured representation that captures attack patterns, asset relationships, and operational contexts within a unified semantic framework.

\paragraph*{STIX 2.1 Extensions and Relationship Modeling}
Grid-STIX extends STIX 2.1's attack pattern objects to incorporate power system-specific tactics, techniques, and procedures classified by grid domain impact (generation, transmission, distribution), target asset types, and operational consequences. The ontology incorporates vulnerability objects that link protocol weaknesses to specific grid components and captures attack progression through relationship objects that model lateral movement paths, privilege escalation vectors, and cascading failure dependencies across grid topology and trust domains.

\paragraph*{Contextual Threat Intelligence}
Grid-STIX integrates operational, environmental, and cyber contexts to provide situational threat intelligence. Attack patterns are linked to operational contexts (peak demand, maintenance windows, emergency conditions), environmental contexts (weather events), and cyber contexts (network topology, trust boundaries) that influence attack feasibility and enable threat prioritization based on current operational conditions.

\section{Design of Grid-STIX}

Grid-STIX employs a modular ontological framework that represents complex cyber-physical relationships while maintaining STIX 2.1 compliance, addressing grid-specific modeling challenges through systematic separation of concerns.
\subsection{Modular Structure}

Grid-STIX employs a modular architecture with eleven modules extending STIX 2.1 domain objects while maintaining semantic consistency. 

\paragraph*{Core Foundation Modules}
The ontology foundation consists of three core modules that establish the primary asset and relationship hierarchies. The \textit{Assets Module} defines base asset classes that extend STIX 2.1 domain objects: \texttt{physical-asset} and \texttt{grid-component} inherit from Infrastructure SDO, \texttt{ot-device} extends Software SDO, enabling STIX-compliant threat intelligence integration. The \textit{Components Module} specializes these base classes into comprehensive asset hierarchies including generation assets (\texttt{distributed-energy-resource}, \texttt{centralized-generation-facility}), distribution infrastructure (\texttt{substation}, \texttt{transformer}, \texttt{transmission-line}), and advanced metering infrastructure.

The \textit{Relationships Module} extends STIX Relationship objects to capture power system-specific connections including power flow (\texttt{feeds-power-to}), protection coordination (\texttt{protects-asset}), and control dependencies (\texttt{controls-relationship}). Union classes enable flexible relationship domain and range specifications across heterogeneous asset types.

\paragraph*{Domain-Specific Extension Modules}
Four domain modules provide specialized functionality for grid cybersecurity applications. The \textit{Attack Patterns Module} extends STIX Attack Pattern objects with power system-specific tactics and techniques, including protocol manipulation, protection system attacks, and cascading failure exploitation. The \textit{Policies Module} models Zero Trust policy constructs through access policies, security policies, and policy enforcement mechanisms that integrate with operational contexts.

The \textit{Events and Observables Module} defines grid-specific cyber observables including telemetry data, alarm events, and protocol traffic patterns, extending STIX Observed Data objects. The \textit{Nuclear Safeguards Module} provides specialized modeling for nuclear energy cybersecurity applications, encompassing nuclear power plant operations, research reactor security, and nuclear fuel cycle facilities with IAEA safeguards integration for comprehensive nuclear energy security modeling.

\paragraph*{Contextual Integration Modules}
Four context modules provide environmental integration: \textit{Operational Contexts} (grid operating conditions, maintenance states), \textit{Environmental Contexts} (weather events, natural disasters), \textit{Cyber Contexts} (network segmentation, trust boundaries for Zero Trust), and \textit{Physical Contexts} (security perimeters, access control zones).

\paragraph*{Vocabulary and Integration Framework}
The \textit{Vocabularies Module} provides controlled vocabularies and open enumerations for grid protocols, equipment types, and operational states, ensuring consistent terminology across domain modules. The \textit{Root Integration Module} imports all modules, providing a single entry point for comprehensive grid cybersecurity modeling.

\paragraph*{URI and Label Conventions}
Grid-STIX follows consistent naming conventions to ensure machine readability and semantic clarity. Class and property URIs employ kebab-case naming (e.g., \texttt{distributed-energy-resource}, \texttt{feeds-power-to}) while \texttt{rdfs:label} values use snake\_case formatting (e.g., \texttt{distributed\_energy\_resource}, \texttt{feeds\_power\_to}). This dual convention supports both human readability and automated code generation while maintaining compatibility with STIX naming patterns.

\subsection{Grid Relationships and Policies}

Grid-STIX models power system operational relationships and Zero Trust policy constructs through specialized relationship objects and policy frameworks that capture both physical grid dependencies and cybersecurity enforcement mechanisms.

\paragraph*{Power System Relationship Modeling}
The Relationships Module defines power system-specific connections that extend STIX relationship semantics to capture grid operational dependencies. Power flow relationships (\texttt{feeds-power-to}, \texttt{generates-power-for}) model electrical energy transfer paths between generation assets, transmission infrastructure, and distribution networks. Protection relationships (\texttt{protects-asset}, \texttt{protects-relationship}) capture protective relay coordination and backup protection schemes essential for cascading failure analysis.

Control and monitoring relationships (\texttt{controls-relationship}, \texttt{monitors-relationship}) model supervisory control dependencies and telemetry data flows between operational technology devices and control centers. Grid synchronization relationships capture frequency response, voltage regulation, and load balancing dependencies across interconnected utility systems.

\paragraph*{Authentication and Policy Integration}
Grid-STIX models multi-factor authentication through relationship objects that capture authentication requirements and authorization relationships across grid infrastructure. Zero Trust policies integrate with grid operational contexts to enable threat-informed access control decisions based on real-time operational state, maintenance windows, and emergency response modes. The policy framework supports multiple enforcement actions coordinated with grid operational procedures to prevent cybersecurity responses from compromising grid stability.

\subsection{Protocol Coverage}

Grid-STIX represents industrial control system protocols through controlled vocabularies and asset bindings that enable standardized modeling of communication patterns, protocol stacks, and interface characteristics across heterogeneous grid infrastructure.

\paragraph*{Protocol Modeling Framework}
The Vocabularies Module defines open vocabulary terms for major protocol families including DNP3, Modbus TCP/RTU, IEC 61850, IEC 60870-5-104, OPC-UA, and IEEE communication standards. Grid assets specify supported protocols through standardized interface properties that capture protocol capabilities and communication characteristics. Protocol-specific relationships capture communication patterns between grid assets, modeling master-slave protocols like DNP3 and publish-subscribe patterns for IEC 61850 GOOSE communication.

\paragraph*{Operational Integration}
Protocol representations integrate with operational contexts to capture how communication patterns change during different grid operating modes. The framework supports threat analysis by enabling systematic identification of communication dependencies, protocol transition points, and cross-domain message flows that influence attack surface characteristics across grid infrastructure.

\section{Tooling and Validation}

Grid-STIX employs robust validation mechanisms and automated tooling to ensure semantic consistency, STIX compliance, and practical usability.

\paragraph*{Validation Pipeline}
The validation framework uses ROBOT~\cite{robotTool} to merge modular ontology files and performs structural validation including class connectivity verification, property consistency checking, and STIX 2.1 compliance verification. Quality assurance integration includes Python formatting, security scanning, and dependency auditing to maintain code and ontology integrity.

\paragraph*{Code Generation and Visualization}
Grid-STIX implements a four-stage generation pipeline that transforms OWL ontologies into STIX-compliant Python classes using Owlready2, intermediate representation optimization, and Jinja2 templating. The framework generates interactive HTML visualizations with grid-specific color schemes and filtering capabilities for communication and ontology exploration. Development workflow integration through Makefile targets enables coordinated validation, generation, and quality assurance cycles.

\section{Use Cases and Intended Deployment}

Grid-STIX enables diverse applications across power system cybersecurity domains through use cases spanning threat modeling, supply chain risk assessment, nuclear energy applications, and Zero Trust enforcement.

\subsection{Threat Modeling and Supply-Chain Risk}

\paragraph*{Substation Attack Progression Scenario}
Consider a coordinated attack targeting a transmission substation through compromised protection relay firmware. Grid-STIX models this scenario through interconnected threat intelligence objects that capture attack progression and cascading impacts. The attack begins with a supply chain compromise where malicious firmware is embedded in protection relays during manufacturing, represented through \texttt{supply-chain-risk} objects linked to specific \texttt{supplier} entities and \texttt{ot-device} components.

The Grid-STIX representation captures the attack pattern (\texttt{firmware-attack-pattern}) with its prerequisite conditions including physical access during manufacturing and target relay specifications. Protection relay assets (\texttt{grid-component}) include supplier provenance, firmware version tracking, and protection coordination relationships (\texttt{protects-asset}) that enable impact assessment across the substation's electrical topology.

When the attack activates, Grid-STIX models the progression through relationship chains that capture how compromised relays can manipulate protection coordination, potentially causing cascading equipment damage through \texttt{affects-operation-of} relationships. The threat model includes temporal dependencies, operational contexts (maintenance windows, peak demand periods), and cross-domain impacts affecting both cyber and physical infrastructure.

\paragraph*{Supply Chain Provenance Analysis}
Grid-STIX enables systematic supply chain risk assessment through comprehensive component provenance tracking and vulnerability inheritance modeling. Supply chain entities include detailed supplier information, country of origin tracking, and component dependency mappings that enable risk aggregation across complex equipment hierarchies.

For distributed energy resource deployments, Grid-STIX models multi-tier supplier relationships where solar inverters include components from multiple vendors, each with distinct security postures and country-of-origin risk profiles. The ontology captures how vulnerabilities in communication modules can affect entire distributed generation systems through \texttt{contains} and \texttt{depends-on} relationships.

Risk assessment scenarios leverage Grid-STIX relationship modeling to identify single points of failure, assess supplier concentration risks, and evaluate the impact of supply chain disruptions on grid operational capabilities. This enables utility operators to prioritize supplier diversification, implement enhanced verification requirements for high-risk components, and develop contingency plans for supply chain compromises.

\paragraph*{Nuclear Energy Security and Safeguards Applications}
Grid-STIX's nuclear safeguards module enables cybersecurity modeling for nuclear energy facilities and fuel cycle operations, addressing unique requirements for nuclear non-proliferation verification and facility protection. Nuclear power plant scenarios leverage Grid-STIX to model relationships between reactor control systems, safety instrumentation, and nuclear material accounting systems. The framework enables threat modeling for nuclear fuel cycle facilities where cyber attacks could compromise safeguards verification or material security, supporting cooperation between national nuclear authorities and international organizations such as the IAEA for standardized threat intelligence sharing while protecting sensitive nuclear technology information.

\paragraph*{Cross-Utility Threat Intelligence Sharing}
Grid-STIX facilitates automated threat intelligence sharing between utility organizations through STIX 2.1-compliant representations that preserve operational context while enabling privacy-preserving information exchange. Grid-STIX's STIX compliance enables integration with existing threat intelligence platforms, automated indicator matching, and coordinated response across the electrical sector.

\subsection{Zero-Trust Enforcement}

Grid-STIX provides the semantic foundation for Zero Trust enforcement in distributed energy resource networks by modeling the essential policy constructs, trust relationships, and operational contexts required for fine-grained authorization decisions in grid environments.

\paragraph*{Zero Trust Implementation}
Grid-STIX enables Zero Trust implementations through modeling of policy decision points, policy enforcement points, and trust brokers with grid-specific operational contexts. The framework captures multi-factor authentication requirements and supports threat-informed access control decisions that consider grid operating conditions, maintenance windows, and emergency response modes. Grid-aware authorization contexts include device authentication history, firmware integrity status, and protection relationships that enable policy expressions incorporating operational knowledge such as restricting high-privilege operations during peak demand periods or implementing automatic quarantine procedures when anomalies are detected.

\paragraph*{Privacy-Preserving Integration}
Grid-STIX enables privacy-preserving threat intelligence sharing through abstraction mechanisms that protect sensitive operational details while enabling collaborative cybersecurity intelligence. Grid-STIX's STIX 2.1 compliance enables integration with existing threat intelligence platforms while providing grid-specific extensions for operational technology environments, supporting complex operational scenarios including technician access controls, emergency override procedures, and cross-utility coordination.

\section{Availability and Community Adoption}

The complete Grid-STIX implementation is available on GitHub at \url{https://github.com/argonne-appres/grid-stix} under the GPLv3 license. The repository provides modular OWL ontology files, build systems, validation tooling, and visualizations. Each release includes versioned ontology bundles enabling stable integration with platforms and research applications.

The Python package generation system produces STIX-compliant classes enabling direct integration with cybersecurity toolchains. Interactive HTML visualizations provide accessible interfaces for exploring Grid-STIX concepts and relationships.

Grid-STIX employs contribution guidelines and issue templates that enable community engagement and ontology evolution. The project follows semantic versioning principles to ensure compatibility across releases while enabling extension of vocabularies, asset classes, and relationship types. Pull request workflows include automated validation and consistency checking to maintain quality.

Contribution pathways support diverse community needs including extensions for emerging grid technologies, new attack pattern modeling, and integration with additional frameworks including MITRE ATT\&CK for ICS. Issue templates guide contributors through ontology enhancement processes while maintaining semantic consistency and STIX compliance.

\section*{Acknowledgments}
This software was developed under U.S. Department of Energy award DE-CR0000049, issued by the Office of Cybersecurity, Energy Security, and Emergency Response (CESER). The prime contractor on this work was Iowa State University, and the ideas herein are influenced by conversations with them. The submitted manuscript has been created by UChicago Argonne, LLC, operator of Argonne National Laboratory. Argonne, a DOE Office of Science laboratory, is operated under Contract No. DE-AC02-06CH11357. The U.S. Government retains for itself, and others acting on its behalf, a paid-up nonexclusive, irrevocable worldwide license in said article to reproduce, prepare derivative works, distribute copies to the public, and perform publicly and display publicly, by or on behalf of the Government.

\bibliographystyle{IEEEtran}
\bibliography{references}

\end{document}